\documentclass[pre,amsmath,amssymb,twocolumn,superscriptaddress,showpacs]{revtex4-1}

\usepackage{amsmath,amssymb}

\usepackage[usenames]{color}

\usepackage{amssymb}

\usepackage{grffile}

\usepackage[pdftex]{graphicx}

\usepackage{amsmath, amstext, amssymb, amsfonts, amsxtra}

\usepackage{textcomp}

\usepackage{xspace}

\def \ell{{d}}

\newcommand{\sutd}{Singapore University of Technology and Design, 8 Somapah Road, 487372 Singapore} 

\newcommand{\majulab}{MajuLab, CNRS-UNS-NUS-NTU International Joint Research Unit, UMI 3654, Singapore}

\begin{document}

\title{Quantum statistics and the performance of engine cycles}

\author{Yuanjian Zheng}

\affiliation{\sutd}


\author{Dario Poletti}

\affiliation{\sutd} 

\affiliation{\majulab}

\begin{abstract}

We study the role of quantum statistics in the performance of Otto cycles. First, we show analytically that the work distributions for bosonic and fermionic working fluids are identical for cycles driven by harmonic trapping potentials. Subsequently, in the case of non-harmonic potentials, we find that the interplay between different energy level spacings and particle statistics strongly affects the performances of the engine cycle. To demonstrate this, we examine three trapping potentials which induce different (single particle) energy level spacings: monotonically decreasing with the level number, monotonically increasing, and the case in which the level spacing does not vary monotonically.       

\end{abstract}

\maketitle

\section{Introduction}
A major technological challenge today is the design of devices at the nanoscale that can efficiently and reliably convert energy between different forms. At such small scales, the thermal and quantum fluctuations of thermodynamic quantities become especially significant, and the stochastic nature plays a dominant role in determining the performance of quantum devices \cite{CampisiTalkner2011,Seifert2012,Kosloff2013, BenentiSaito2013,GelbwaserKlimovskyKurizki2015}. 

More recently, major theoretical developments have also been made specifically towards the understanding and implementation of quantum thermodynamics. These include for instance, advances in fluctuations theorems \cite{CampisiTalkner2011, Seifert2012, HanggiTalkner2015}, and on the use of external control protocols to produce adiabatic dynamics in finite time \cite{DemirplakRice2005, Berry2009, DengGong2013, PalmeroMuga, TorronteguiMuga2013, CampbellFazio2014, DeffnerDelCampo2014, DelCampoPaternostro2014}. 

At the same time, properties of non-equilibrium work distributions have also been investigated experimentally in classical systems \cite{BlickleBechinger2006, GomezSolanoBechinger2015}, culminating with the realization of a colloidal micro-scale stochastic heat engine\cite{BlickeBechinger2012}, and quantum systems \cite{AnKim2015}. Due to the high degree of tunability and control, ultra-cold ion setups have also positioned themselves as promising candidates for experimental realizations of quantum heat engine cycles \cite{AbahLutz2012}. In addition, a proposal for the realization of an Otto cycle within a solid state set-up was recently studied \cite{CampisiFazio2015}.  

More recently, it has also been shown numerically, that the exchange symmetry of particles do play a significant role in the work distribution of quantum many-body systems \cite{GongQuan2014}. As such, here we address the pertinent issue of many-body statistics in quantum heat engines. In particular, we elucidate the role of exchange statistics in the performance of quantum cycles by providing an intuitive understanding of the interesting connection between exchange statistics and the geometry of the trapping potential.

We consider a quantum equivalent of the Otto-cycle \cite{QuanNori2007} in which the working fluid is a quantum gas of $N$ indistinguishable and non-interacting particles. The collective system undergoes a cyclic sequence of adiabatic and iso-parametric processes, where the heat received by external reservoirs can be in part, converted into useful work, while the rest is transferred to a cold reservoir. Important characteristics and figures of merit of the engine cycle's performances are the probability distributions of both its net work output and its efficiency. In this work we study how the statistics, either bosonic or fermionic, of the gas' particles affects the engine's performance. We show that this is strongly dependent on the single particle energy level structure of the system and indicate strategies to improve the performance of the engine cycle by a careful choice of both the trapping potential and the type of gas. It is only when considered in conjunction, that trap geometry \cite{ZhengPoletti2014} and particle exchange statistics \cite{GongQuan2014} can be utilized to optimize the performance of quantum heat engines.

\begin{figure}
\includegraphics[width=\columnwidth]{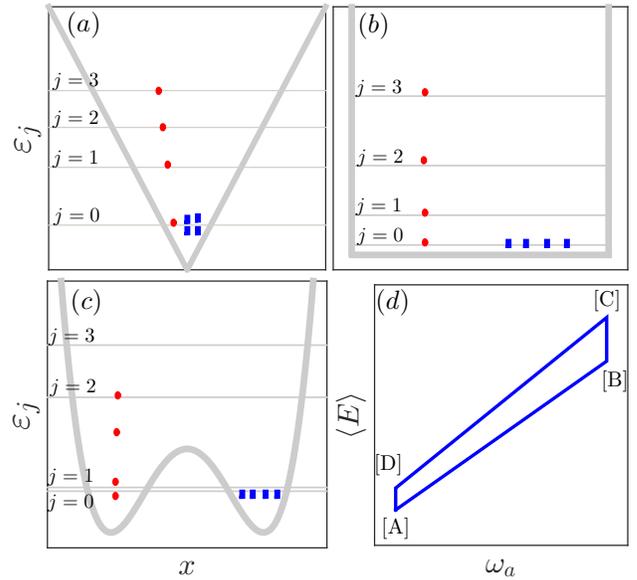}
\caption{(Color online) (a-c) Different potential wells with corresponding single particle energy spectrum $\varepsilon_j$. The dark blue (light red) squares represent the occupation of energy levels at 0 temperature for the case of $4$ bosonic (fermionic) atoms. (a) triangular potential ($a=1$), (b) infinite square well potential ($a=\infty$) and (c) double well potential. The horizontal thin grey lines indicate the position of the $j$-th single particle energy level. (d) Schematics of a quantum Otto cycle in an average energy $\langle E\rangle$ vs trapping parameter $\omega_a$.} \label{fig:1} 
\end{figure}

We consider non-interacting particles in different trapping potentials: with the energy level spacings either constant (harmonic potential), monotonically decreasing (triangular potential, see Fig.\ref{fig:1}(a)), increasing (square potential, see Fig.\ref{fig:1}(b)) or change non-monotonically (double well, see Fig.\ref{fig:1}(c)). To gain some preliminary intuitive insights to the problem, in Fig.\ref{fig:1} we show how $4$ atoms are distributed in the respective potentials at temperature $T=0$ depending on the given symmetry. We note that the energy difference between the ground state and the first excited state can be very different depending on whether the gas is bosonic or fermionic in nature and on whether the gap between energy levels is increasing (e.g. square well) or decreasing (e.g. triangular potential). In Fig.\ref{fig:1}(a) we observe that for bosons (blue squares) the gap is larger than for that of the fermions (red circles); Whereas in the case of a potential in which the energy level distance is monotonically increasing (square well in Fig.\ref{fig:1}(b), the gap will be larger for the fermions and smaller for that of the bosons. For a potential in which the distances between energy levels do not feature a monotonous behavior (double well in Fig.\ref{fig:1}(c)), the excitation gap not only depends on the particle statistics, but also on the number of atoms. 

This paper is organized as follows: In section \ref{sec:2}, we provide general relations and results for quantum Otto cycles with scaling potentials; In section \ref{sec:3}, we show our main results and lastly, in section \ref{sec:4} we draw our conclusions.  

\section{The Otto Cycle in scaling potentials \label{sec:2}}   
In classical thermodynamics, the Otto cycle is composed of two processes in which only heat is exchanged with the environment, and two processes in which only work is exchanged (and no heat is transferred). The former is usually referred to as an isochoric process for a classical ideal gas. Analogously, a quantum Otto cycle is composed of two quantum adiabatic processes, represented in Fig.\ref{fig:1}(d) by the steps $[A]\rightarrow [B]$ and $[C]\rightarrow [D]$, as well as two iso-parametric processes, represented in Fig.\ref{fig:1}(d) by the steps $[B]\rightarrow [C]$ and $[D]\rightarrow [A]$, in which the Hamiltonian of the system is kept constant and the environment is (weakly) coupled to the system \cite{weakcoupling}. Similar to their classical counterparts, no heat is exchanged with the working fluid in quantum adiabatic processes, while only heat is exchanged in an iso-parametric process. $T_{[A]}$ corresponds to the lowest temperature in the cycle while $T_{[C]}$ is the largest. As we are interested in the governing principles of quantum thermodynamic cycles involving non-interacting many-body quantum gases, we will only study quasi-static cycles, leaving the cycles with finite power for a future study.

To gain deeper analytical insights, we also focus on Hamiltonians with scaling properties \cite{QuanNori2007, Jarzynski2013, DeffnerDelCampo2014, ZhengPoletti2014}, or more generally on driving protocols in which the energy level structure after a unitary process is simply multiplied by a scale-factor. The harmonic oscillator is the prototypical example of such scaling Hamiltonians, as the energy levels are simply proportional to the trapping frequency.

In a more general sense, for every potential of the form $V(x_i)=f(x_i/\lambda)/\lambda^2 $ (where $\lambda$ is a control parameter with units of length and $x_i$ is the position coordinate of the $i$-th atom), the Hamiltonian $H$ has a scaling energy spectrum. In fact    
\begin{align} 
H&=\sum_i\left[-\frac{\hbar^2}{2m}\frac{\partial^2}{\partial x_i^2} + \frac{1}{\lambda^2} f\left(\frac {x_i} \lambda \right)\right]\nonumber\\ 
&=\frac 1 {\lambda^2} \sum_i \left[-\frac{\hbar^2}{2m}\frac{\partial^2}{\partial X_i^2} + f\left( X_i \right) \right]
\end{align}
where $X_i=x_i/\lambda$ is a dimensionless variable \cite{DeffnerDelCampo2014}.  
This applies to all power-law potentials of the form $V(x_i)=\frac{1}{2}m(\omega_a |x_i|)^a$ since 
\begin{align} 
H&=\sum_i-\frac{\hbar^2}{2m}\frac{\partial^2}{\partial x_i^2} + \frac{m}{2}\left(\omega_a|x_i|\right)^a \nonumber\\
&=\sum_i\zeta_a(\omega_a)\left(-\frac{1}{2}\frac{\partial^2}{\partial X_i^2} + \frac{1}{2}|X_i|^a\right)
\end{align}
where $\zeta_a(\omega_a)=\left[\left(\hbar\omega_a\right)^{2a}/m^{a-2}\right]^{\frac{1}{a+2}}$ is the energy scale \cite{Jarzynski2013, ZhengPoletti2014}. It is thus now clear that that the energy levels are simply scaled by the prefactor $\zeta_a(\omega_a)$. For instance, the ratio of the energy of the $j$-th level for Hamiltonian parameter $\omega_a'$ (i.e. $\varepsilon_j(\omega_a')$) divided by that for parameter $\omega_a$ (i.e. $\varepsilon_j(\omega_a)$) does not depend on $j$ and it is given by 
\begin{align} 
\frac{\varepsilon_j(\omega_a')}{\varepsilon_j(\omega_a)}=\frac{\zeta_a(\omega_a')}{\zeta_a(\omega_a)}=\nu \label{eq:nu}
\end{align}  
where $\nu$ is a parameter that only depends on $\omega_a$ and $\omega'_a$. In the following we will use the lighter notation $\zeta_a(\omega_a')=\zeta_a'$ and $\zeta_a(\omega_a)=\zeta_a$.  

In addition, a more general polynomial like the double-well potential may also be made to possess a scaling property in its energy levels, albeit in the presence of certain restrictions. For example, the double well potential 
\begin{equation}
V(x_i)=\frac{m}2  \left(-\omega_2^2x_i^2 +\omega_4^4x_i^4\right)
\end{equation}
has a scaling Hamiltonian
\begin{equation}
H=\zeta_2\left(-\frac{1}{2}\frac{\partial^2}{\partial X_i^2}-\frac{1}{2}X_i^2+\frac{\hbar}{m}\frac{\omega_4^4}{\omega_2^3}X_i^4\right)
\end{equation}
so long as the anharmonic parameter $\gamma \equiv \hbar \omega_4^4/m\omega_2^3$ remains fixed throughout the Hamiltonian evolution.

\subsection{Mean Work $\langle W_p \rangle $ } 
Let us consider a process from Hamiltonian parameter $\omega_a$ to $\omega_a'$. The respective single particle $j$-th energy eigenvalues will be $\varepsilon_{j}$ and $\varepsilon'_{j}$.   
Let $n_{j}$ represent the number of particles in the $j$-th single energy eigenstate. 

A single $N-$body state is thus given by the ordered set of occupation numbers $\{n_j\}$. A process is performed quasi-statically by changing the trap parameter adiabatically, and the corresponding $N-$body microstate at the end of the compression is then given by $\{ m_k\}$. For a single realization of the protocol $p$ the work is given by a two-times energy measurement \cite{TalknerHanggi2007, CampisiTalkner2011}  
\begin{equation}
W_p=E'_{\{n_j\}}- E_{\{m_k\}}
\end{equation}
The energy $E_{\{m_k\}}$ (resp. $E'_{\{n_j\}}$) is given by the sum over the energy $\varepsilon_{l}$ (resp. $\varepsilon'_{l}$) of the $l$-th single particle eigenstate of the Hamiltonian (e.g. $E_{\{m_k\}} = \sum_{k} \varepsilon_{k} m_k$). Hence, the ensemble average of work done for the process between two trap parameters is given by:
\begin{align}
\langle W_p \rangle&=\sum_{\{n_j\}}\sum_{\{m_k\}}(E'_{\{m_k\}}-E_{\{n_j\}})\delta_{{\{m_k\}},{\{n_j\}}}P_{\{{n_j}\}}\nonumber\\
&=\sum_{\{n_j\}}\left(E'_{\{n_j\}}-E_{\{n_j\}}\right)P_{\{{n_j}\}}
\end{align}
where the Kronecker delta $\delta_{{\{m_k\}},{\{n_j\}}}$ is due to the adiabaticity of the process. The average $\langle\dots\rangle$ is computed over the intial probability distribution $P_{\{n_j\}}$ which, at this stage, can be general. Because of the scaling property of the Hamiltonians we consider, the ratio of initial to final energy of each single particle eigenstate $(\varepsilon'_{j} / \varepsilon_{j})=\nu$ (see Eq.(\ref{eq:nu}) and \cite{DeffnerDelCampo2014, ZhengPoletti2014}). 
This results in the work of an adiabatic process to be  
\begin{equation}
\langle W_p \rangle =\left(\nu-1\right) \langle E \rangle  \label{eq:aw}    
\end{equation}
where 
\begin{equation}
\langle E \rangle = \sum_{\{n_j\}}\sum_{j}\varepsilon_j n_jP_{\{n_j\}}               
\end{equation}
is the energy of the system at the beginning of the process.  

\subsection{Work Standard Deviation $\sigma_{W_p}$ }
The standard deviation of the work probability distribution function of a process $p$, $\sigma_{W_p}=\sqrt{\langle W_p^2\rangle - \langle W_p\rangle^2}$, is also important to characterize a process as it quantifies the consistency and reliability of the output. For an adiabatic unitary process in a scaling potential we have       
\begin{align} 
\langle W_p^2 \rangle &=\sum_{\{n_j\}}\sum_{\{m_k\}}(E'_{\{m_k\}}-E_{\{n_j\}})^2\delta_{{\{m_k\}}{\{n_j\}}}P_{\{{n_j}\}} \nonumber\\ 
&=\left(\nu-1\right)^2\sum_{\{n_j\}}\left(\sum_{j} \varepsilon_{j} \; n_j\right)^2 P_{\{n_j\}}  \nonumber\\ 
&=\left(\nu-1\right)^2\langle E^2 \rangle \label{eq:sd}
\end{align}   
It follows that the scaled standard deviation of work, defined as the standard deviation of the work distribution divided by its mean work is given by 
\begin{equation}
\frac{\sigma_{W_p}}{\langle W_p \rangle}\equiv \sqrt{\frac{\langle W_p^2 \rangle}{\langle W_p\rangle^2}-1} = \sqrt{\frac{\langle E^2 \rangle}{\langle E \rangle^2}-1} \label{eq:sd2}
\end{equation}
which is independent of $\nu$, i.e. independent of whether (i) the process is a compression or an expansion, and (ii) the extent of the process (provided it is quantum adiabatic).

\subsection{Net work output of a cycle: average and variance}

The ensemble moments of the cyclic work distribution is derived from the work distribution of the constitutive single adibatic processes. The cycle work distribution is composed of two statistically independent work distributions on the compression and expansion arms of the cycle because of the nature of the ideal quantum Otto cycle. The combined work distribution is thus given by the product distribution of two statistically independent processes    
\begin{equation}
P(W)=\delta[W- (W_c+W_e)]P(W_c)P(W_e)
\end{equation} 
where $W_c$ ($W_e$) is the work in the compression (expansion) process. For the cycle depicted in Fig.\ref{fig:1}(d) between parameters $\omega_a$ and $\omega'_a$ the net work output is given by 
\begin{equation} 
\langle W \rangle_{\alpha}=(\nu-1)\langle E_{[A]} \rangle_{\alpha} + \left(\frac 1 \nu -1\right) \langle E_{[C]} \rangle_{\alpha} \label{eq:network}    
\end{equation} 
where here again, $\nu=\zeta'_a/\zeta_a$ and $\langle E_{[V]} \rangle_{\alpha}$ is the average energy at the vertex ${[V]}$ of the cycle and $\langle\dots\rangle_{\alpha}$ means average over the thermal canonical state for fermions ($\alpha=F$) or bosons ($\alpha=B$).

Consequentially, the mean and variance of the cycle are also just the linear sum of the mean and variance of the individual processes. The scaled work fluctuations are thus given by
\begin{equation}\label{eq:scaled_work}
\frac{\sigma_{W_{\alpha}}}{\langle W \rangle_{\alpha}}=\frac{\sqrt{(\Delta W_{c_{\alpha}})^2+(\Delta W_{e_{\alpha}})^2}}{\langle W_{c} \rangle_{\alpha}+\langle W_{e} \rangle_{\alpha}}
\end{equation}
In addition, we also note that if the adiabatic processes of an Otto cycle are peformed within scaling potentials then the efficiency is a non-stochastic function of $\nu$ \cite{ZhengPoletti2014} independent of the quantum statistics of the working fluid. At each realization of the cycle  
\begin{align}
\eta&=-\frac{ W }{ Q_{[B]\rightarrow [C]} }\nonumber\\ 
&=\frac{(1-1/\nu) E_{[C],\{m_k\}} + (1-\nu) E_{[A],\{n_j\}} }{E_{[C],\{m_k\}}-\nu E_{[A],\{n_j\}}}\nonumber\\
&=1-\frac 1 \nu \label{eq:effic}
\end{align}
where indeed $\nu>1$ and $ Q_{[B]\rightarrow [C]} $ is the heat transferred into the system between states $[B]$ and $[C]$. In Eq.(\ref{eq:effic}) $E_{[V],\{m_k\}}$ represents the energy of the eigenstate $\{m_k\}$ at vertex $[V]$ of the cycle.

\section{Cycles performance for bosonic and fermionic gases \label{sec:3}}

The average work output and its standard deviation can thus be computed from Eq.(\ref{eq:aw}) and (\ref{eq:sd}). 
The average energy and variance of a canonical thermal state is computed using $\langle E \rangle_{\alpha} =- \partial \left(\ln{Z^{\alpha}_N}\right)/\partial \beta$ and $\sigma^2_{E_{\alpha}} = \partial^2 \left(\ln{Z^{\alpha}_N}\right)/\partial^2 \beta$ where $\beta=1/k_B T$ is the inverse temperature and $k_B$ is the Boltzmann constant. 
To obtain accurate results for low number of particles and low temperatures, we use the canonical parition function $Z^{\alpha}_N$ for $N$ (fermionic or bosonic) atoms \cite{grandcanonical}. It can be computed using the following recursion relation for non-degenerate spectra   
\begin{equation}
Z^{\alpha}_N(\beta)=\frac{1}{N}\sum_{n=1}^N (-1)^\gamma Z^{\alpha}_1(n \beta)Z^{\alpha}_{N-n}(\beta) \label{eq:BosFerPart}   
\end{equation}
where $\gamma=2n$ for bosons ($\alpha=B$) \cite {WilkensWeiss1997,WeissWilkens1997} and $\gamma=n+1$ for fermions ($\alpha=F$) \cite{Kashcheyevs2011} and where $Z^{\alpha}_1(\beta) = \sum_j e^{-\beta\epsilon_j}$ is the single particle partition function (which is obviously independent of the value of ${\alpha}$). Note also that $Z^{\alpha}_0=1$. Eq.(\ref{eq:BosFerPart}) (derived in detail in the Appendix A), will be used to compute $Z^{\alpha}_N(\beta)$ numerically. For the harmonic case, exact analytical results will be described in the following section (III-A).

\begin{figure}
\includegraphics[width=\columnwidth]{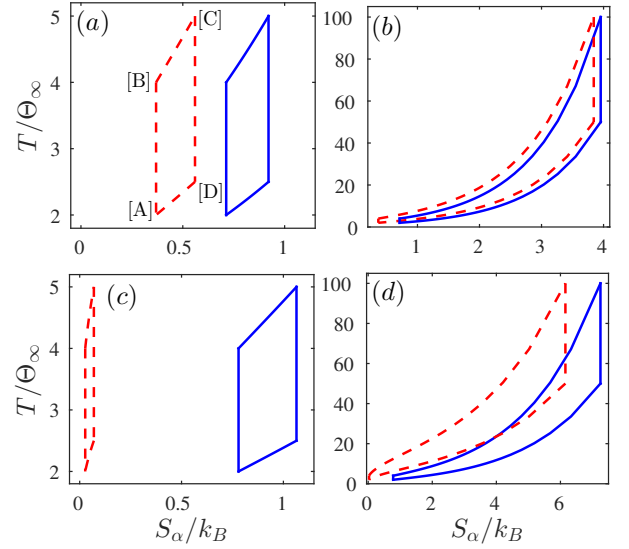}
\caption{(Color online) Temperature $T$ vs entropy $S_{\alpha}$ diagram for an Otto cycle in an infinite square well potential ($a=\infty$) with fermion (red-dashed line) or bosons (blue-continuous line) as working substances. The trapping parameter varies between $\zeta_\infty$ and $\zeta'_{\infty}=2\zeta_\infty$ while the minimum temperature $T_{[A]}= 2\; \Theta_{\infty}$. The particle number $N$ and the maximum temperature $T_{[C]}$ takes the following values: (a) $N=2$ and $T_{[C]}=5\; \Theta_{\infty} $, (b) $N=2$ and $T_{[C]}=100\; \Theta_{\infty}$, (c) $N=5$ and $T_{[C]}=5\; \Theta_{\infty}$, (d) $N=5$ and $T_{[C]}=100\; \Theta_{\infty}$. } \label{fig:fig2} 
\end{figure}

\begin{figure}
\includegraphics[width=\columnwidth]{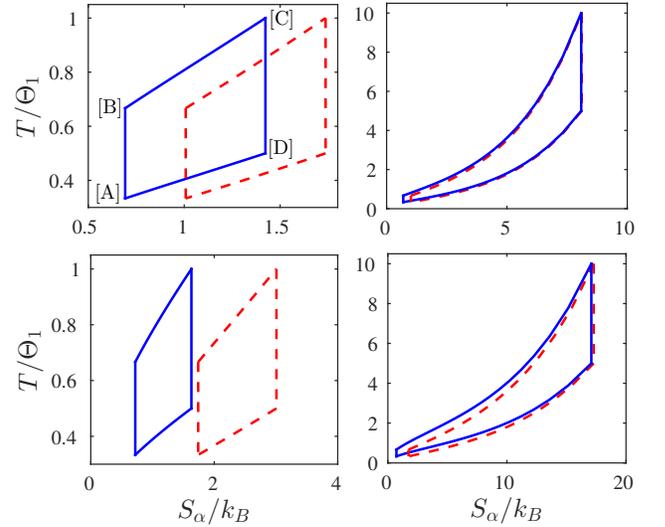}
\caption{(Color online) Temperature $T$ vs entropy $S_{\alpha}$ diagram for an Otto cycle in a triangular potential ($a=1$) with fermions (red-dashed line) or bosons (blue-continuous line) as working substances. The trapping parameter varies between $\zeta_{1}$ and $\zeta'_{1}=2\zeta_1$ while the minimum temperature $T_{[A]}= \Theta_1/3$. The particle number $N$ and the maximum temperature $T_{[C]}$ take the following values: (a) $N=2$ and $T_{[C]}= \Theta_{1}$, (b) $N=2$ and $T_{[C]}=10 \;\Theta_{1}$, (c) $N=5$ and $T_{[C]}=\Theta_{1}$, (d) $N=5$ and $T_{[C]}=10\; \Theta_{1}$. } \label{fig:fig3}      
\end{figure}

\subsection{Harmonic potential} 
For a simple harmonic oscillator of frequency $\omega_2$, it is possible to compute the partition function with the use of the $q-$shifted factorials $(z;q)_n\equiv \prod_{n=1}^N(1-zq^{n-1})$. The partition function for $N$ bosons is given by    
\begin{equation}
Z^{B}_N=\frac{e^{-N\beta\hbar\omega_2/2}}{(q;q)_N}
\end{equation}
and for $N$ fermions \cite{Kashcheyevs2011} by 
\begin{equation}
Z^{F}_N=\frac{{e^{-N^2\beta\hbar\omega_2/2}}}{(q;q)_N}
\end{equation}
where $q=e^{\beta\hbar\omega_2}$ (see Appendix B for a detailed derivation). This results in average energies, $\langle E \rangle_{\alpha}$, given by 
\begin{equation}
\langle E \rangle_B= \frac{\hbar\omega_2 N}{2}+\sum^N_{n=1} \frac{\hbar\omega_2 n}{ e^{\beta \hbar\omega_2 n}-1}\label{eq:bospart}
\end{equation}
and 
\begin{equation}
\langle E \rangle_F= \langle E \rangle_B + \frac{\hbar\omega_2}2 N\left(N-1\right).   \label{eq:ferpart}
\end{equation}
All higher moments of the energy are identical for bosons and fermions. In particular, the standard deviation for bosons $\sigma_{E_B}$ is equal to that of fermions $\sigma_{E_F}$ 
\begin{equation} 
\sigma_{E_B}=\sigma_{E_F}=\sum^N_{n=1} \frac{(\hbar\omega_2 n)^2 e^{\beta \hbar\omega_2 n}}{(e^{\beta \hbar\omega_2 n}-1)^2}
\end{equation}      
The constant mean energy difference between fermions and bosons ensures that the average work transfer is always greater for that of fermions, $\langle W_{p} \rangle_F > \langle W_{p} \rangle_B$, while the rescaled standard deviation is instead, always larger for bosons $\frac{\sigma_{W_{p_F}}}{\langle W_{p} \rangle_F} < \frac{\sigma_{W_{p_B}}}{\langle W_{p} \rangle_B}$ \cite{equality}. This of course assumes that the comparisons are between given initial canonical states of the same temperatures and identical quantum adiabatic protocols in a harmonic potential. 

For the work output of the full cycle, with Hamiltonian parameter that changes between $\omega_2$ and $\omega_2'$, using Eq.(\ref{eq:network}) one can also easily show that
\begin{align}     
\langle W \rangle_F &= \langle W \rangle_B + \left[ (\nu-1) \omega_2 + \left(\frac 1 \nu -1 \right) \omega'_2 \right] \Delta  \nonumber \\ 
&= \langle W \rangle_B + \left[ \left(\frac{\omega'_2}{\omega_2}-1\right) \omega_2 + \left(\frac {\omega_2} {\omega'_2} -1 \right) \omega'_2 \right] \Delta \nonumber \\
&= \langle W \rangle_B   
\end{align}    
where $\Delta=\frac{\hbar}{2}N(N-1)$. 

Hence for a reversible quantum Otto cycle driven by harmonic potentials with non-interacting particles, all moments of work distribution for the cycle are the same, regardless of the working fluid being bosonic or fermionic.

\subsection{Triangular and square well potentials}

\begin{figure}
\includegraphics[width=0.9\columnwidth]{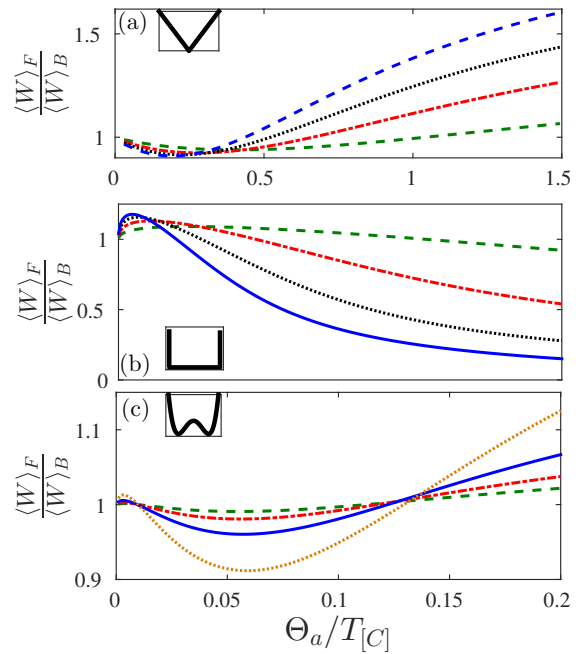}
\caption{(Color online) Fermionic to bosonic ratio of the mean work output of an Otto cycle vs. high temperature $T_{[C]}$ for different particle numbers in the following potentials: (a) Triangular potential $(a=1)$, (b) Infinite square well potential $(a=\infty)$ and (c) Double well potential ($a=2, \gamma=0.01$). In all potentials, $\zeta'_a=2\zeta_a$.
In (a) $T_{[A]}= \Theta_1/3$, while in (b) and (c) $T_{[A]}=2\Theta_a$. Different numbers of atoms are represented by the different (colored) lines: $N=2$ for the dashed green line, $N=3$ for the dot-dashed red line, $N=4$ for the dotted black line and $N=5$ for the continuous blue line and $N=10$ for the dotted brown line in (c). The insets depict the potentials used.} \label{fig:fig4}    
\end{figure}

\begin{figure}
\includegraphics[width=0.9\columnwidth]{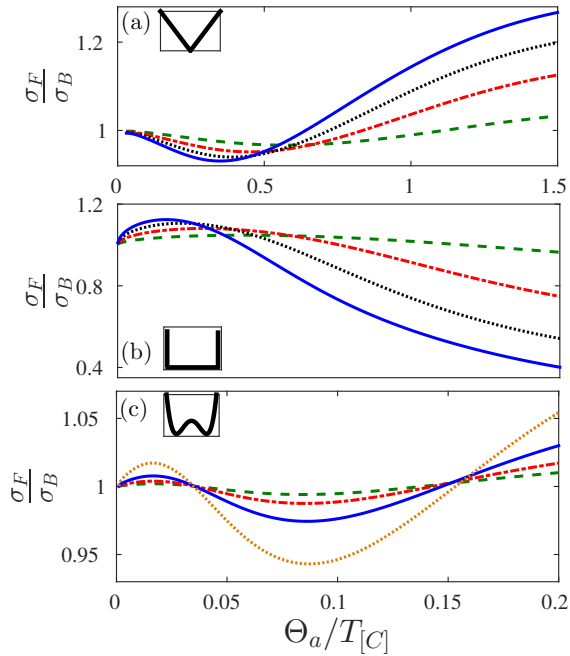}
\caption{(Color online) Fermionic to bosonic ratio of the standard deviation in the work distribution of an Otto cycle vs. high temperatures $T_{[C]}$ for different particle numbers. Values for the cold temperature $T_{[A]}$, particle number $N$ and Hamiltonian parameter values $\omega_{a}$ are as given in Fig.\ref{fig:fig4}. The insets depict the potentials used.} \label{fig:fig5} 
\end{figure}

The previous results for the harmonic oscillator are a direct consequence of the constant energy level spacings. For non-harmonic potentials, where the level spacing is not constant, the results will be very different. For instance, from Fig.\ref{fig:1}(a) it is clear that, at zero temperature, a fermionic system requires less energy to be excited compared to a bosonic one (because the gap between consecutive energy levels monotonically decreases, hence the energy gap from the last filled state to the first unfilled one is larger for bosons than for fermions). The converse is true for a square well potential in which the energy level spacing monotonically increases, see Fig.\ref{fig:1}(b). As we will show later, this has strong consequences on the performance of an engine cycle. 

To describe this in a clear and pictorial manner, see Figs.\ref{fig:fig2} and \ref{fig:fig3}, we use a representation of a cycle typical of classical thermodynamics, the temperature vs entropy $T$-$S$ diagram (where $T$ is divided by a scale of temperature given by $\Theta_a=\zeta_a/k_B$ while in the cycle the Hamiltonian parameter varies between $\omega_a$ and $\omega_a'>\omega_a$). To compute the entropy $S_{\alpha}$ we use the thermodynamic relation $F_{\alpha}=\langle E \rangle_{\alpha}-TS_{\alpha}$, where the free energy $F_{\alpha}=-k_B T\ln(Z^{\alpha}_N)$. In this diagram, the area within the cycle represent both the net heat and net work exchanges. 

Using Eq.(\ref{eq:BosFerPart}) we are able to compute $Z^{\alpha}_N$ for different particle numbers and cycle parameters.           

We consider only cases in which the coldest temperature $T_{[A]}$ is cold enough to see the combined effect of both energy level spacings and particle statistics \cite{lowtemperature}. We then consider different cases in which the value of the highest temperature $T_{[C]}$, varies from (relatively) cold to hot, and also with working fluids that consist of different number of particles. For low $T_{[C]}$ we expect large differences in the distribution for bosonic or fermionic gases.  
In Fig.\ref{fig:fig2} we depict the case of an infinite square well, while in Fig.\ref{fig:fig3} we study the case of a triangular potential. In both figures the fermionic cycle is represented with dashed red lines and the bosonic cycle by continuous blue lines. Whether a bosonic or a fermionic working fluid produces more average work output depends on the difference between the adiabatic work in the cold (compression from $[A]\to [B]$) and in the hot arm (expansion from $[C]\to [D]$).   

As clearly shown in Fig.\ref{fig:fig2}(c), when the number of particles are large and both temperatures $T_{[A]}$ and $T_{[C]}$ are cold enough, there can be remarkable differences between the work output of a fermionic or a bosonic system. The area enclosed in the fermionic cycle is much smaller. In fact, for fermions the energy level separation is too large for entropy to change significantly. A similar scenario, although in the opposite direction, emerges for a triangular potential as depicted in Fig.\ref{fig:fig3}(c). In this case the energy level spacing is larger for the bosons and the entropy will be lower (and less heat is absorbed). For small number of atoms and large enough highest temperatures, the ratio of the net work output $\langle W\rangle_F/\langle W\rangle_B$ tends to $1$, see Figs.\ref{fig:fig2}(b) and \ref{fig:fig3}(b). 

The case of two atoms and cold temperatures are depicted in Figs.\ref{fig:fig2}-\ref{fig:fig3}(a). In this scenario the number of atoms is too small to result in a marked difference of the cycles due to the quantum statistics. Cycles for larger number of atoms and hotter temperatures are shown in Figs.\ref{fig:fig2}-\ref{fig:fig3}(d). It should be noted that for the triangular potential, Fig.\ref{fig:fig3}(d), almost no difference can be noted between bosons and fermions in the high temperature regime. However, in the cold temperature portion of the cycle, important and large differences in the amount of entropy exchanged can be noticed. In particular, the entropy is lower for the species experiencing the largest gap (bosons in this case). 

To gain deeper understanding we revert to Figs.\ref{fig:fig4}-\ref{fig:fig5}(a-b). The triangular and the square well potential have qualitatively opposite behaviors. When the highest temperature $T_{[C]}$ is small, the total entropy and the entropy change for the bosons in the triangular potential (respectively for the fermions in the square well potential) are much smaller than that of the fermions (respectively bosons), see Figs.\ref{fig:fig2}(c) and \ref{fig:fig3}(c). 
It is easier to understand the qualitative behavior for large $T_{[C]}$ by considering the work exchanges. In this case the ratio of the work exchanges $\langle W_{[C]\rightarrow [D]} \rangle_F / \langle W_{[C]\rightarrow [D]} \rangle_B = \langle E_{[C]} \rangle_F/\langle E_{[C]} \rangle_B$ and thus it will eventually converge towards $1$ as the temperature increases. However the difference in the work processes $\langle W_{[C]\rightarrow [D]} \rangle_F - \langle W_{[C]\rightarrow [D]} \rangle_B$ behaves differently depending on the type of potential. It increases indefinitely for the square well, while it goes to $0$ for the triangular potential. Hence, for the square well potential the ratio $\langle W \rangle_{F} / \langle W \rangle_{B}$ will tend to $1$ from above, while for the triangular potential it will tend to $1$ from below \cite{frombelow}. The same analysis also explains the qualitatively similar behavior of the variance of the work probability distribution, Fig.\ref{fig:fig5}. For this quantity, however, the crossings of the line $\sigma_F/\sigma_B=1$ occur at a different values of $T_{[C]}$.

\subsection{Double well potentials} 

The double well (Fig.\ref{fig:1}(c)), is a typical example of a system in which the energy level spacing does not vary monotonically. Hence, we will use this to study the interplay of geometry and particle statistics in the performance of the engine cycle. As shown earlier in section \ref{sec:2}, the double well potential can have a scaling property so long as $\gamma \equiv \hbar \omega_4^4/m\omega_2^3$ is a constant. For ease of analytical and numerical analysis, we will focus on this case. As shown in Fig.\ref{fig:fig4}(c) the fermionic to bosonic ratio of the work output $\langle W_F \rangle/\langle W_B \rangle$ has a pronounced oscillatory behavior compared to potentials with monotonically changing energy level structure as the (inverse) temperature of the hot reservoir $\beta_h$. 

The same behavior is expected and observed also for the ratio of the standard deviation of the work done for a fermionic over a bosonic working fluid $\sigma_F/\sigma_B$, see Fig.\ref{fig:fig5}(c).

\section{Conclusions \label{sec:4}}   

In this work we have studied how quantum statistics of particles in a working fluid can affect the performance of an engine cycle. We have shown that the interplay of statistics with energy level structure of a system has remarkable and non-trivial effects on the engine cycle's output. We have analyzed systems with monotonicallly increasing (square well potential) and decreasing (triangular potential) energy level structure, which possess qualitatively opposite behaviors, both in the low and high temperature regimes. For potentials with a non-monotonic energy level spacing (for example, a double well potential), the engine cycle's peformance is characterized by the presence of oscillations in the fermionic to bosonic ratio of a given figure of merit, and the properties of these oscillations are closely associated to the underlying energy level structure of the Hamiltonian, the particle number and the temperature regimes. As a final remark, systems driven in finite time, and many-body interactions will be the subject of future studies. 

We aknowledge insightful discussions with C. Kollath at an early stage of the work. We are also grateful to U. Bissbort, J. Gong, C. Guo, P. H\"anggi, R. Tan, and G. Xiao for fruitful conversations. We are supported by SUTD Start-up grant EPD2012-045 and by the SUTD-MIT International Design Centre (IDC).\\

\begin{appendix} 

\section{Derivation of recursive relation for the canonical partition function  \label{app:a}}  

Here, we derive Eq.(\ref{eq:BosFerPart}) following the argument as presented in  \cite{Kashcheyevs2011}. To do so we start from the grand canonical partition function $\Xi_{\alpha}$, where ${\alpha}=B$ or $F$, is the subscript that denotes bosonic and fermionic statistics respectively. The grand partition function can be written as  
\begin{align} 
\Xi_{\alpha}=1+\sum_{N=1}^\infty Z^{\alpha}_{N}z^N  \label{eq:grandcano}
\end{align}     
where $Z^{\alpha}_{N}$ is the canonical partition function for $N$ atoms. In equation (\ref{eq:grandcano}) $z=e^{\beta\mu}$ is commonly known as the fugacity, where $\mu$ is the chemical potential associated to the ensemble. 

It is also possible to write the logarithm of $\Xi_{\alpha}$ as 
\begin{align} 
\ln\left(\Xi_{\alpha}\right)=\sum_{N=1}^{\infty} \frac{\kappa_{{\alpha},N}}{N!}z^N \label{eq:loggrandcano}
\end{align}     
where $\kappa_{{\alpha},N}$ in general, assumes different functional forms depending on the given type of particle statistics considered. By exponentiating (\ref{eq:loggrandcano}) and using the Fa\`a di Bruno's formula, which states that the power series expansion of the exponential of a polynomial $\sum_N a_N x^N/N!$ is 
\begin{equation}
\exp{\left(\sum_N a_N x^N/N!\right)}=\sum_{N=0}^{\infty} B(a_1, ... a_N) x^N/N!  
\end{equation} 
we can derive
\begin{align} 
\Xi_{\alpha}=e^{\left(\sum_{n=1}^{\infty} \frac{\kappa_{{\alpha},_N}}{N!}z^N\right)} = \sum_{N=0}^{\infty} B_N\left(\kappa_{{\alpha},1}, ..., \kappa_{{\alpha},_N}\right) \frac{z^N}{N!}  \label{eq:polycano}   
\end{align} 
where $B_N$ is a Bell polynomial. Next, by equating corresponding terms of the polynomial expansion in (\ref{eq:polycano}) and (\ref{eq:grandcano}), we get 
\begin{align} 
Z^{\alpha}_{N} = \frac{B_N(\kappa_{{\alpha},1},...\kappa_{{\alpha},_N})}{N!}.  \label{eq:kappacano}   
\end{align}   
Note that up to this point in the derivation, we have not made any distinction between bosonic or fermionic statistics. The grand partition function for the fermions can be written as a formal power series 
\begin{align} 
\ln(\Xi_{F}) &= \sum_k \ln\left(1+z e^{-\beta\varepsilon_k}\right) \nonumber\\ 
&= \sum_k \sum_N (-1)^{N+1} (N-1)!\; e^{-N\beta\varepsilon_k} \frac{z^N}{N!} \nonumber\\ 
&= \sum_N (-1)^{N+1} (N-1)!\; \left(\sum_k e^{-N\beta\varepsilon_k}\right) \frac{z^N}{N!} \nonumber\\ 
& =  \sum_N (-1)^{N+1} (N-1)!\; Z_{1}(N\beta) \frac{z^N}{N!}
\label{eq:kappaFer}   
\end{align}   
where we have used the expansion of $\ln(1+x)$ and where $Z_1(n\beta)$ is the single particle canonical partition function with inverse temperature $n\beta$ (of course $Z_1=Z^F_{1}=Z^B_{1}$). Finally, using equations (\ref{eq:kappacano}) and (\ref{eq:kappaFer}) together with the identity for Bell polynomials:
\begin{align}
B_n(\kappa_1, ...\kappa_n) = \kappa_n + \sum_{m=1}^{n-1} \left(\begin{array}{c} n-1 \\ m-1 \end{array} \right) \kappa_m B(\kappa_1, ...\kappa_{n-m}) \label{eq:Bell}
\end{align}
we derive Eq.(\ref{eq:BosFerPart}) for fermions.   
Similarly for bosons  
\begin{align} 
\ln(\Xi_{B}) &= -\sum_k \ln\left(1-z e^{-\beta\varepsilon_k}\right) \nonumber\\ 
&= -\sum_k \sum_n (-1)^{2n+1} (n-1)!\; e^{-n\beta\varepsilon_k} \frac{z^n}{n!} \nonumber\\ 
&=  \sum_n (n-1)!\; Z_{1}(n\beta) \frac{z^n}{n!} \label{eq:kappaBos}   
\end{align}   
and again using Eq.(\ref{eq:kappacano}), (\ref{eq:Bell}) and (\ref{eq:kappaBos}) yields Eq.(\ref{eq:BosFerPart}) for bosons.\\

\section{Partition function for the simple harmonic oscillator \label{app:b}}  

For the simple harmonic oscillator, the Hamiltonian parameter $a=2$. Hence the energy of the system will scale with $\hbar\omega_2$. 
Following the derivation of \cite{Kashcheyevs2011}, we begin by writing the grand canonical partition function for bosons in terms of a q-shifted factorial:
\begin{align}
\Omega_B(z)=\prod^\infty_{k=0}\frac{1}{1-q^k \tilde{z}}=\frac{1}{(\tilde{z};q)_{\infty}}
\end{align}
Here, $q = e^{-\beta\hbar\omega_2}$, $\tilde{z}=ze^{-\beta\hbar\omega_2/2 }$ where  $z=e^{\beta\mu}$ is as before, the fugacity, and $(\tilde{z};q)_{\infty}$ is a q-shifted factorial. By using the $q$-analog binomial theorem:
\begin{align}
\frac{1}{(\tilde{z};q)_{\infty}}=\sum^{\infty}_{n=0}\frac{\tilde{z}^n}{(q;q)_{\infty}}
\end{align}
where the $q$-shifted factorial reduces to simply:
\begin{equation}
(q;q)_N =\prod^N_{x=1} (1-q^x)
\end{equation}
Notice that we can also  write $\Omega(z)$ as a formal power series involving the canonical partition function $Z^{\alpha}_N$
\begin{equation}
\Omega_{\alpha}(z)=1+\sum_{n=0}^{\infty}Z^{\alpha}_Nz^n
\end{equation}
And thus by comparison of the two equivalent series, we are able to write down the canonical partition function for the bosons:
\begin{equation}
Z^{B}_N=\frac{e^{-N\beta\hbar\omega_2/2}}{(q;q)_N}
\end{equation}
Similarly for the fermions,
\begin{align}
\Omega_F(z)=(-\tilde{z};q)_{\infty} 
\end{align}
and using the identity 
\begin{align}
(\tilde{z},q)_{\infty}=\sum_{N=0}^{\infty} \frac{(-1)^N q^{N(N-1)/2}\;\tilde{z}^N}{(q;q)_N} 
\end{align}
we can write
\begin{equation}
Z^{F}_N=\frac{{e^{-N^2\beta\hbar\omega_2/2}}}{(q;q)_N}
\end{equation}
\end{appendix}


\begin{thebibliography}{10}
\bibitem{grandcanonical} The use of the grand-canonical partition function would not give accurate results in the regimes of particle numbers and temperatures we study. Moreover the canonical partition function allows us easy access to energy fluctuations for fixed particle number.
\bibitem{equality} Note that for the Harmonic oscillator $\langle W \rangle_F$ and $\langle W \rangle_B$ approach the same value in the high temperature limit.
\bibitem{lowtemperature} If $T_{[A]}$ was too high, the difference between the fermionic and bosonic statistics will be much reduced.
\bibitem{frombelow} In this case it plays particular importance the fact that at cold temperatures the work done in the cold arm is, in modulus, larger for the fermions than for the bosons. Hence, while the difference in work becomes negligible in one process, in the other process there is more work exchange for the fermions. The combination of these two effects results in a net work which is lower for the fermions.
\end{thebibliography}
\end{document}